\documentclass[aps,twocolumn,groupedaddress,showpacs,showkeys,nofootinbib,amsfonts,eqsecnum,floatfix]{revtex4}

\usepackage[utf8]{inputenc}
\usepackage{amsmath}
\usepackage{amsfonts}
\usepackage{amssymb}
\usepackage{pstricks}
\usepackage{pst-plot}

\begin{document}

%%%%%%%%%%%%%%%%%%%%%%%%%%%%%%%%%%%%
%%%%%%%%%%%%%%%%%%%%%%%%%%%%%%%%%%%%
%%%%         Title              %%%%
%%%%%%%%%%%%%%%%%%%%%%%%%%%%%%%%%%%%
%%%%%%%%%%%%%%%%%%%%%%%%%%%%%%%%%%%%
\title{The Heavy Quark Free-Energy at $T \le T_c$ in AdS/QCD}
%%%%%%%%%%%%%%%%%%%%%%%%%%%%%%%%%%%%
%%%%%%%%%%%%%%%%%%%%%%%%%%%%%%%%%%%%
%%%%          Authors           %%%%
%%%%%%%%%%%%%%%%%%%%%%%%%%%%%%%%%%%%
%%%%%%%%%%%%%%%%%%%%%%%%%%%%%%%%%%%%
\author{K. Veshgini}
\email{K.Veschgini@tphys.uni-heidelberg.de}
\author{E. Meg\'{\i}as}
\email{emegias@tphys.uni-heidelberg.de}
\author{H.J. Pirner}
\email{pir@tphys.uni-heidelberg.de}

\affiliation{Institut für Theoretische Physik der Universität Heidelberg, Heidelberg, Germany}

\author{J.Nian}
\email{nian@sissa.it}

\affiliation{International School for Advanced Studies, Trieste, Italy}

\date{\today}

%%%%%%%%%%%%%%%%%%%%%%%%%%%%%%%%%%%%
%%%%%%%%%%%%%%%%%%%%%%%%%%%%%%%%%%%%
%%%%          Abstract          %%%%
%%%%%%%%%%%%%%%%%%%%%%%%%%%%%%%%%%%%
%%%%%%%%%%%%%%%%%%%%%%%%%%%%%%%%%%%%
\begin{abstract}
Starting with the modified AdS/QCD metric developed in Ref.~\cite{Pirner:2009gr} we use the Nambu-Goto action to obtain the free energy of a quark-antiquark pair at $T<T_\mathrm{c}$, for which we show that the effective string tension goes to zero at $T_\mathrm{c}=154\, \textrm{MeV}$.
\end{abstract}



\maketitle
%%%%%%%%%%%%%%%%%%%%%%%%%%%%%%%%%%%%
%%%%%%%%%%%%%%%%%%%%%%%%%%%%%%%%%%%%
%%%%        Introduction        %%%%
%%%%%%%%%%%%%%%%%%%%%%%%%%%%%%%%%%%%
%%%%%%%%%%%%%%%%%%%%%%%%%%%%%%%%%%%%
\section{Introduction}
\label{sec:intro}
%%%%%%%%%%%%%%%%%%%%%%%%%%%%%%%%%%%%
%       Fig. Wilson-Loop           %
%        --------------            %
%       (  )        (  )           %
%        --------------            %
%%%%%%%%%%%%%%%%%%%%%%%%%%%%%%%%%%%%
\begin{figure}[t]
\psset{xunit=3cm,yunit=2cm,arrowscale=1.5}
\begin{pspicture}(-1.4,-1.3)(1.5,1.3)
	\psellipticarc{-<}(-1,0)(.3,1){0}{360}
	\psellipticarc{>-}(0.7,0)(.3,1){0}{90}
	\psellipticarc[linestyle=dotted]{-}(0.7,0)(.3,1){90}{270}
	\psellipticarc{-}(0.7,0)(.3,1){270}{360}
	\uput[180](1,0){$q$}
	\uput[-45](-.7,0){$\overline{q}$}
	\psline{->}(-1.1,0)(0.8,0)
	\uput[90](0.8,0){$x$}

	\uput[-90](-1,0){$-\frac{d}{2}$}
	\psdot(-1,0)
	\uput[-90](0.7,0){$\frac{d}{2}$}
	\psdot(0.7,0)

	\psellipticarc{->}(0.65,0)(.3,1){35}{60}
	\uput[0](0.7,0.7){$\tau$}
	\psline{->}(-1,0)(-.79,0.71)
	\uput[90](-.95,.36){$R$}
	\thinlines
	\psline[linearc=10](-1,1)(-.15,.8)(.7,1)
	\psline[linearc=10](-1,-1)(-0.15,-0.8)(.7,-1)
\end{pspicture}
\caption{Wilson loops in Euclidean time with periodicity $\beta=2\pi R$.}
\label{fig:WL} 
\end{figure}
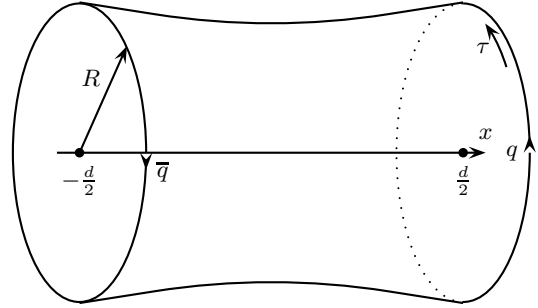

As shown in Ref.~\cite{McLerran:1981pb} the free energy of a static,
infinitely massive quark-antiquark pair $F$ is given
by

\begin{equation} \label{fqq1} 
e^{-\beta F}=\left<L(\vec{r_q}) L^\dagger(\vec{r_{\overline{q}}}) \right>\,,
\end{equation}
where $L(\vec{r})$ is the Wilson-Line

\begin{equation}
L(\vec{r})=\frac{1}{N}\mathrm{tr}\,\mathcal{T}\exp\left[i\int\limits_{0}^{\beta} d\tau \hat{A}^0(\vec{r},\tau) \right]\,,
\end{equation}
$\beta=1/T$ is the inverse temperature, $\hat{A}^0$ is the gluon field
in the fundamental representation and $\tau$ denotes imaginary time.
According to the holographic dictionary \cite{Witten:1998qj} the right
hand side of equation (\ref{fqq1}) is equal to the string partition
function on the $\mathrm{AdS}_5$ space with the integration contours
on the boundary of $\mathrm{AdS}_5$. In saddle-point approximation:

\begin{equation}
 e^{-\beta F}=\left<L(\vec{r_q})
 L^\dagger(r_{\overline{q}}) \right> \approx e^{-S_\mathrm{NG}}\,,
\end{equation}
where $S_\mathrm{NG}$ is the Nambu-Goto action
\begin{equation}
S_\mathrm{NG}=\frac{1}{2\pi l_{s}^2}\int d^2\xi\sqrt{\mathrm{det} h_{ab}}\,,
\end{equation}
with the induced worldsheet metric
\begin{equation}
h_{ab}=G_{\mu\nu}\frac{\partial X^\mu}{\partial \xi^a}\frac{\partial X^\nu}{\partial \xi^b}\,.
\end{equation}
As described in
reference \cite{Nian:2009mw}, due to the symmetry of the problem, we
may set up a cylindrical coordinate system in five-dimensional
Euclidean space. Then the five coordinates are:

\begin{itemize}
\item t - time
\item z - the bulk coordinate (extra 5th dimension)
\item x, r, $\phi$ - three spatial coordinates
\end{itemize}

We parameterize an element of the surface connecting the two Wilson loops with $d\xi_1=r d\phi$
and $d\xi_2=dx$, set $t=0$ and reinterpret $d\tau\equiv r d\phi$ as
Euclidean time.  We use the modified metric $G_{\mu\nu}$ developed in reference
\cite{Pirner:2009gr}  effectively for $N_c=3$ and $N_f=4$.
This metric has been further analysed as a possible solution of 
5-d gravity in Ref.~\cite{Galow:2009kw}
\begin{eqnarray}
ds^{2}_{Eucl}&=&\frac{h(z)L^2}{z^2}(r^2d\phi^2+dr^2+dx^2+dz^2)  \,,\\
h(z)&=&\frac{\log(\epsilon)}{\log\left((\Lambda z)^2+\epsilon\right)} \,, \\
\Lambda &=& L^{-1}=264\mathrm{MeV} \,,\\
\epsilon &=& \frac{l_{s}^2}{L^2}=0.48 \,.
\end{eqnarray}
The Nambu-Goto action determines the string surface with $l_s$ as string length:

\begin{eqnarray}
S_\mathrm{NG}&=&\frac{1}{2\pi l_{s}^2}\int\limits_{0}^{2\pi}d\phi
\int\limits_{-d/2}^{d/2} dx \frac{L^2
h(z)}{z^2}r\sqrt{1+(z')^2+(r')^2}\nonumber \\ \label{SNG}
&=&\frac{1}{\epsilon}\int\limits_{-d/2}^{d/2} dx
\frac{h(z)}{z^2}r\sqrt{1+(z')^2+(r')^2} \\ \label{Lag}
&=:&\frac{1}{\epsilon}\int\limits_{-d/2}^{d/2} dx\,
\mathcal{L}[z(x),z'(x),r(x),r'(x)]\,,
\end{eqnarray}

The quark and antiquark are separated by a distance $d$ along the
x-axis. The two Polyakov loops are approximated by Wilson loops of
radius $R=\beta/2\pi$. $\mathcal{L}$ is the Lagrangian density and the
prime ($'$) denotes the derivative with respect to $x$. The configuration in shown in figure~\ref{fig:WL}.

%%%%%%%%%%%%%%%%%%%%%%%%%%%%%%%%%%%%
%%%%%%%%%%%%%%%%%%%%%%%%%%%%%%%%%%%%
%%%   Euler-Lagrange Equations  %%%%
%%%%%%%%%%%%%%%%%%%%%%%%%%%%%%%%%%%%
%%%%%%%%%%%%%%%%%%%%%%%%%%%%%%%%%%%%

\section{Euler-Lagrange Equations}
\label{sec:euler-eq}

Since the Lagrangian in (\ref{Lag}) does not depend on $x$ explicitly,
it is invariant under variations $x\rightarrow x+\delta x$. It follows
from Noether's theorem that there exists a conserved quantity $k$
given by

\begin{equation} \label{k}
k=\frac{h(z)\cdot r}{z^2}\frac{1}{\sqrt{1+(z')^2+(r')^2}}\,.
\end{equation}

The Euler-Lagrange equations corresponding to the Nambu-Goto action
(\ref{SNG}) can be simplified with Eq.~(\ref{k}) (see
Ref.~\cite{Nian:2009mw}):

\begin{equation} \label{EL}
\begin{split} 
r''&-\frac{h^2(z)\cdot r}{k^2 z^4}=0 \,,\\
z''&-\frac{h(z)\cdot r^2\cdot\left(z\partial_z h(z)-2h(z)\right)}{k^2z^5}=0\,.
\end{split} 
\end{equation}

The boundary conditions are

\begin{equation} \label{BC}
\begin{split}
r\left(\pm d/2\right)&=R=\frac{\beta}{2\pi}=\frac{1}{2\pi T}\,,\\
z\left(\pm d/2\right)&=0\,.
\end{split}
\end{equation}

%%%%%%%%%%%%%%%%%%%%%%%%%%%%%%%%%%%%
%%%%%%%%%%%%%%%%%%%%%%%%%%%%%%%%%%%%
%%%%    Numerical Solutions     %%%%
%%%%%%%%%%%%%%%%%%%%%%%%%%%%%%%%%%%%
%%%%%%%%%%%%%%%%%%%%%%%%%%%%%%%%%%%%

\section{Numerical Solutions}
\label{sec:num-sol}
%%%%%%%%%%%%%%%%%%%%%%%%%%%%%%%%%%%%
%       Fig. ELSolReg              %
%       ^                          %
%       |***                       %
%       |***                       %
%       |*****                     %
%       |**********                %
%       |--------------->          %
%%%%%%%%%%%%%%%%%%%%%%%%%%%%%%%%%%%%
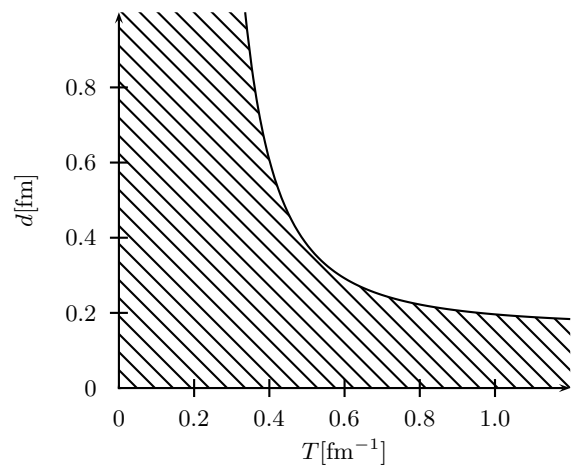
\begin{figure}[hb!]
\psset{xunit=5cm,yunit=5cm}
\begin{pspicture}(-.25,-.15)(1.2,1.0)
\psaxes[Dx=0.2,Dy=0.2]{->}(1.2,1.0)
\pscustom[fillstyle=vlines,fillcolor=black,linestyle=none]{%
	\psplot[plotstyle=curve]{0.335652}{1.2}{0.0253303 x 0.162 sub dup mul div 0.16 add }
	\psline(1.2,0.18351)(1.2,0)
	\psline(1.2,0)(0,0)
	\psline(0,0)(0,1)
	\psline(0,1)(0.335652,1)
}
\psplot[plotstyle=curve]{0.335652}{1.2}{0.0253303 x 0.162 sub dup mul div 0.16 add }
\uput[-90](0.6,-.1){$T[\mathrm{fm}^{-1}]$}\uput[180]{90}(-.18,0.5){$d[\mathrm{fm}]$}
\end{pspicture}
\caption{The Euler-Lagrange equations (\ref{EL}) have only solutions for temperature $T$ and quark-antiquark separation $d$ within the shaded area.}
\label{fig:ELSolReg} 
\end{figure}

Analysis of symmetry (cf. Fig.~\ref{fig:WL}) gives for the first derivatives:

\begin{equation}\label{IC}
\begin{split}
r'(0)&=0\,, \\
z'(0)&=0\,.
\end{split}
\end{equation}

But the conditions Eqs.~(\ref{BC}) and (\ref{IC}) are not given at the same point. It is more convenient to find conditions for the functions and their derivatives
at the same point.
Analysis of Eq.~(\ref{EL}) shows that $r'$ and $z'$ must
diverge near the boundary $x\rightarrow\pm\frac{d}{2}$. In order to obtain some
stable numerical solution, we have studied 
the behavior of $r(x)$ and $z(x)$ near the boundary, cf. Ref.~\cite{Nian:2009mw}.
Numerically, a small cutoff $v$ is applied, then
$r(-d/2+v)$, $z(-d/2+v)$, $r'(-d/2+v)$ and $z'(-d/2+v)$ can be calculated asymptotically
and used as initial conditions. We do not prescribe
the value of $d$. For a fixed value of $R$, we give an
arbitrary value to $k>0$, then calculate $r,z$
and consequently the Nambu-Goto action $\mathrm{S}_{\mathrm{NG}}$ for this value
of $k$. By changing the value of $k$ we obtain the Nambu-Goto action as a
function of $k$, denoted by $\mathrm{S}_{\mathrm{NG}}(k)$. We can express $d$
as a function of $k$, and determine the distance $d$ associated
with the Nambu-Goto action $\mathrm{S}_{\mathrm{NG}}(k)$.
In principle, the constructed numerical solutions $r'(x)$ and $z'(x)$ can vanish at different
points due to the small difference between our
asymptotic solution and the real solution. We adjust the initial
conditions at the point $v$ keeping the value of $k$ fixed in such a
way that $r'(0)$ and $z'(0)$ vanish at the same point. 
The Nambu-Goto action is given by
\begin{equation}
\begin{split}
S_\mathrm{NG}&=\frac{2}{\epsilon}\int_{-d/2}^{0} dx \frac{h(z)\cdot r}{z^2}\sqrt{1+\left(z'\right)^2+\left(r'\right)^2} \\
&=\,\frac{2}{\epsilon}\int_{-d/2}^{-d/2+v} dx \frac{h\left(z_a\right)\cdot
r_a}{z_a^2}\sqrt{1+\left(z_a'\right)^2+\left(r_a'\right)^2}\\
&+\frac{2}{\epsilon}
\int_{-d/2+v}^{0} dx\frac{h\left(z_n\right)\cdot
r_n}{z_n^2}\sqrt{1+\left(z_n'\right)^2+\left(r_n'\right)^2},
\end{split}
\end{equation}
where the subscript ``a'' denotes ``asymptotic solution'' near $x=-d/2$, while ``n'' means
``numerical solution''. In the last expression, the first integral is divergent
at $x=-d/2$. But, as we have the explicit form of the asymptotic solutions
$r_a(x)$ and $z_a(x)$, we can expand the first integrand into power series near
$x=-d/2$, and remove the divergent terms. To compensate this removal, we should add
the antiderivative of the divergent terms at $x=-d/2+v$. This way, we obtain the
regularized value of $\mathrm{S}_{\mathrm{NG}}$.

Integrating the Euler-Lagrange equations (\ref{EL}) for a wide range of initial values suggests that solutions exist only for a specific range of temperature $T$ and quark-antiquark separation $d$ (Fig.~\ref{fig:ELSolReg}).

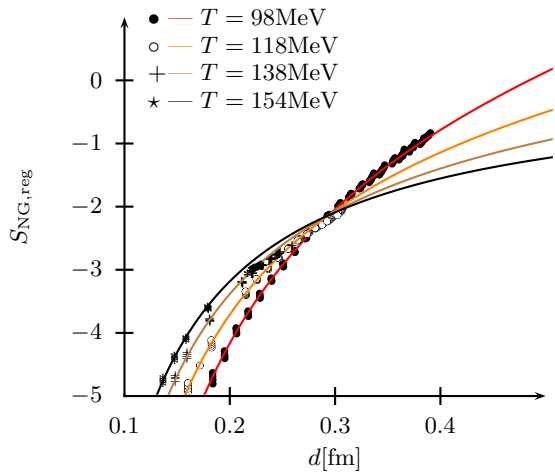
\begin{figure}
\psset{xunit=14cm,yunit=0.84cm}
\begin{pspicture}(0.0,-6)(0.50,1.0)
\psaxes[Dx=0.1,Dy=1,Ox=0.1,Oy=-5]{->}(0.1,-5)(0.1,-5)(0.5,1.0)

\uput[-90](0.3,-5.6){$d[\mathrm{fm}]$}
\uput[180]{90}(.025,-2){$S_\mathrm{NG,reg}$}
\newcommand{\sngfit}[1]{5.8352 #1 div -7.46424 add x mul -0.4786 x #1 mul div add}

\psclip{\psframe[linestyle=none](0,-5)(.5,1)}

\readdata{\data}{./data/T05.txt}
\dataplot[plotstyle=dots,dotsize=.1,dotstyle=*]{\data0}
\psplot[plotstyle=curve,linecolor=red]{.1}{.5}{\sngfit{.5}}

\readdata{\data}{./data/T06.txt}
\dataplot[plotstyle=dots,dotsize=.1,dotstyle=o]{\data0}
\psplot[plotstyle=curve,linecolor=orange]{.1}{.5}{\sngfit{.6}}

\readdata{\data}{./data/T07.txt}
\dataplot[plotstyle=dots,dotsize=.1,dotstyle=+]{\data0}
\psplot[plotstyle=curve,linecolor=brown]{.1}{.5}{\sngfit{.7}}

\readdata{\data}{./data/T08.txt}
\dataplot[plotstyle=dots,dotsize=.1,dotstyle=asterisk]{\data0}
\psplot[plotstyle=curve]{.1}{.5}{\sngfit{.78}}

\endpsclip
\put(1.6,0){\parbox{5cm}{
\begin{tabbing}
$\bullet$ \=\textcolor{red}{---} \=$T=98\mathrm{MeV}$\\
$\circ$\>\textcolor{orange}{---}\>$T=118\mathrm{MeV}$\\
$+$\>\textcolor{brown}{---}\>$T=138\mathrm{MeV}$\\
$\star$\>\textcolor{black}{---}\>$T=154\mathrm{MeV}$
\end{tabbing}
}}
\end{pspicture}
\caption{Regularized Nambu-Goto action as a function of quark-antiquark separation $d$ for different temperatures $T$.}
\label{fig:SNG} 
\end{figure}

Figure \ref{fig:SNG} defines the regularized Nambu-Goto action $S_\mathrm{NG,reg}$ for several temperatures $T$ as a function of quark-antiquark separation $d$. A fit to the numerical calculations gives
\begin{equation}\label{Sfit}
S_\mathrm{NG,reg}^\mathrm{fit}=\frac{-0.48}{T d}+d\left(\frac{-7.46}{\mathrm{fm}}+\frac{5.84}{T \mathrm{fm}^2} \right)\,.
\end{equation}
%%%%%%%%%%%%%%%%%%%%%%%%%%%%%%%%%%%%
%       Fig.  SNG                  %
%       ^           *******        %
%   SNG |       ****               %
%       |     **                   %
%       |    *                     %
%       |   *                      %
%       |--------------------> d   %
%%%%%%%%%%%%%%%%%%%%%%%%%%%%%%%%%%%%

%%%%%%%%%%%%%%%%%%%%%%%%%%%%%%%%%%%%
%%%%%%%%%%%%%%%%%%%%%%%%%%%%%%%%%%%%
%%%   Thermodynamic Quantities   %%%
%%%%%%%%%%%%%%%%%%%%%%%%%%%%%%%%%%%%
%%%%%%%%%%%%%%%%%%%%%%%%%%%%%%%%%%%%
\section{Thermodynamic Quantities}
\label{sec:Therm}

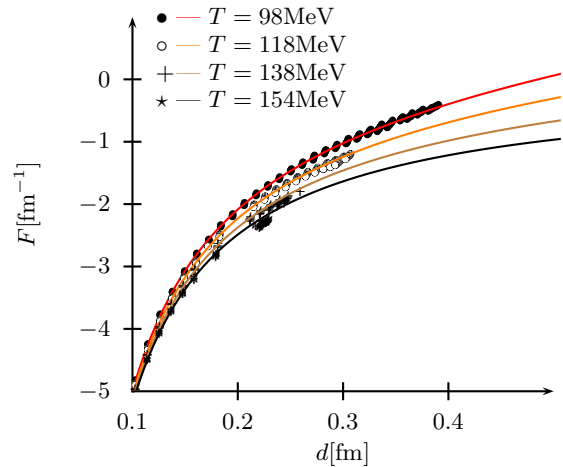
\begin{figure}
\psset{xunit=14cm,yunit=.83cm}
\begin{pspicture}(0.0,-6)(0.50,1.0)
\psaxes[Dx=0.1,Dy=1,Ox=0.1,Oy=-5]{->}(0.1,-5)(0.1,-5)(0.5,1.0)

\uput[-90](0.3,-5.6){$d[\mathrm{fm}]$}
\uput[180]{90}(.025,-2){$F[\mathrm{fm}^{-1}]$}
\newcommand{\sngfit}[1]{5.8352 #1 div -7.46424 add x mul -0.4786 x #1 mul div add #1 mul}

\psclip{\psframe[linestyle=none](0,-5)(.5,1)}

\readdata{\data}{./data/FT05.txt}
\dataplot[plotstyle=dots,dotsize=.1,dotstyle=*]{\data0}
\psplot[plotstyle=curve,linecolor=red]{.05}{.5}{\sngfit{.5}}

\readdata{\data}{./data/FT06.txt}
\dataplot[plotstyle=dots,dotsize=.1,dotstyle=o]{\data0}
\psplot[plotstyle=curve,linecolor=orange]{.05}{.5}{\sngfit{.6}}

\readdata{\data}{./data/FT07.txt}
\dataplot[plotstyle=dots,dotsize=.1,dotstyle=+]{\data0}
\psplot[plotstyle=curve,linecolor=brown]{.05}{.5}{\sngfit{.7}}

\readdata{\data}{./data/FT08.txt}
\dataplot[plotstyle=dots,dotsize=.1,dotstyle=asterisk]{\data0}
\psplot[plotstyle=curve]{.05}{.5}{\sngfit{.78}}

\endpsclip
\put(1.6,0){\parbox{5cm}{
\begin{tabbing}
$\bullet$ \=\textcolor{red}{---} \=$T=98\mathrm{MeV}$\\
$\circ$\>\textcolor{orange}{---}\>$T=118\mathrm{MeV}$\\
$+$\>\textcolor{brown}{---}\>$T=138\mathrm{MeV}$\\
$\star$\>\textcolor{black}{---}\>$T=154\mathrm{MeV}$
\end{tabbing}
}}
\end{pspicture}
\caption{Free energy as a function of quark-antiquark separation $d$ for different temperatures $T$.}
\label{fig:F} 
\end{figure}

Using $S_\mathrm{NG,reg}^\mathrm{fit}$ it is easy to calculate the free energy of the $Q\bar{Q}$-system as a function of the $Q\bar{Q}$ separation:
\begin{equation}\label{F}
F=T\cdot S_\mathrm{NG,reg}^\mathrm{fit}= \frac{-0.48}{d}+d\left(\frac{-7.46}{\mathrm{fm}}T+\frac{5.84}{\mathrm{fm}^2} \right)\,.
\end{equation}
Fig.~\ref{fig:F} shows the free energy. One can recognize  the flattening of $F$ when increasing temperature~$T$ for large ditances~$d$.
The term linear in $d$ yields the effective string tension:
\begin{equation}\label{sigmaeff}
\sigma_\mathrm{effective}=\frac{-7.46}{\mathrm{fm}}T+\frac{5.84}{\mathrm{fm}^2}\, .
\end{equation}
%%%%%%%%%%%%%%%%%%%%%%%%%%%%%%%%%%%%
%       Fig.  F                    %
%       ^           *******        %
%     F |       ****               %
%       |     **                   %
%       |    *                     %
%       |  *                       %
%       |--------------------> d   %
%%%%%%%%%%%%%%%%%%%%%%%%%%%%%%%%%%%%
The entropy writes as
\begin{equation}
S=-\frac{\partial F}{\partial T}=\frac{7.46}{\mathrm{fm}}d\,,
\end{equation} 
and does not depend on temperature. Such an entropy is well known for strong coupling QCD on a 3 dimensional lattice, where $S = \ell \cdot \log\left( 2D-1 \right)$, with $D=3$ and $\ell$ is the length of the random path in lattice units connecting the quark and antiquark~\cite{MeyerOrtmanns:1996ea}. Inner energy and string tension also do not depend on temperature and are given by
\begin{equation}
E=F+T S=\frac{-0.48}{d}+\frac{5.84}{\mathrm{fm}^2}d\,,
\end{equation}
and
\begin{equation}
\sigma=\frac{5.84}{\mathrm{fm}^2}=\sigma_\mathrm{effective}(T=0)\,,
\end{equation}
respectively.

%%%%%%%%%%%%%%%%%%%%%%%%%%%%%%%%%%%%
%%%%%%%%%%%%%%%%%%%%%%%%%%%%%%%%%%%%
% Confinement and Phase Transition %
%%%%%%%%%%%%%%%%%%%%%%%%%%%%%%%%%%%%
%%%%%%%%%%%%%%%%%%%%%%%%%%%%%%%%%%%%

\section{Confinement and Phase Transition}
\label{sec:conf}

The free energy Eq.~(\ref{F}) contains a linear term and a Coulomb-like term, which is absent in strong coupling lattice QCD. The linear term provides confinement: when it becomes zero there will be no confinement. From Eq.~(\ref{sigmaeff}) we can estimate the critical temperature for the confinement/deconfinement  phase transition, which is roughly $T_\mathrm{c}=154\,\mathrm{MeV}$. This value is in agreement with lattice QCD results, i.e. $T_\mathrm{c}^\mathrm{lattice}(N_c= 3, N_f = 3)=155\pm10\,\mathrm{MeV}$ \cite{Yagi:2005yb}, but one must admit that we only solve a pure gluon theory without dynamical quarks, where, however, the input $T=0$ potential has been fitted for $N_f=4$.
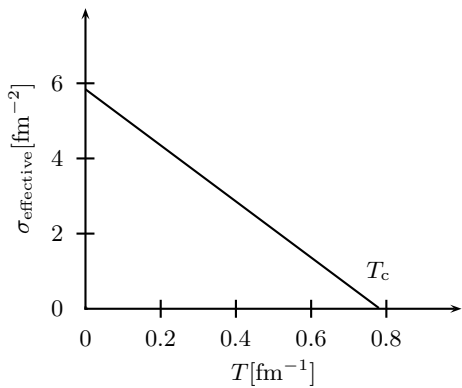
\begin{figure}
\psset{xunit=5cm,yunit=.5cm}
\begin{pspicture}(-.25,-.8cm)(1.0,8.0)
\psaxes[Dx=0.2,Dy=2]{->}(1.0,8.0)
\psplot[plotstyle=curve]{0}{0.78}{-7.46 x mul 5.84 add }
\rput(.78,.5cm){$T_\mathrm{c}$}
\uput[-90](.5,-.5cm){$T[\mathrm{fm}^{-1}]$}
\uput[180]{90}(-.5cm,4){$\sigma_\mathrm{effective}[\mathrm{fm}^{-2}]$}
\end{pspicture}
\caption{Effective string tension as a function of temperature. The phase transition happens when the effective string tension vanishes.}
\label{fig:tension} 
\end{figure}

\section{Conclusion}
\label{sec:conclusion}

We have shown that the modified metric of AdS/QCD proposed in Ref.~\cite{Nian:2009mw} can also be applied to the heavy quark potential. Since we are not using the black hole metric this theory is restricted to the $T<T_\mathrm{c}$ regime. We found that the modified $\mathrm{AdS}_5$-metric produces confinement and the short distance Coulombic behavior in this region. In previous works on loop-loop correlators \cite{Shoshi:2002in,Shoshi:2002rd}, these two features had to be added by hand, whereas here they follow from one action. We also can determine $T_c$ by demanding that the effective string tension vanishes.

Because of the singularity in the metric at $z_\mathrm{IR}=\frac{1}{\Lambda}\sqrt{1-\epsilon}/\approx0.54\mathrm{fm}$ the Euler-Lagrange equations~(\ref{EL}) have only solution for a very limited range of boundary condition. In particular for large $T$ they yield solutions only for very small $q\overline{q}$ separations making the fit~(\ref{Sfit}) more hypothetical.

%\begin{acknowledgments}

%\end{acknowledgments}

%\appendix
%
%\section{Heavy QQ potential}
%\label{app:heavyqq}

%%%%%%%%%%%%%%%%%%%%%%%%%%%%%%%%%%%%
%%%%%%%%%%%%%%%%%%%%%%%%%%%%%%%%%%%%
%%%         Bibliography         %%%
%%%%%%%%%%%%%%%%%%%%%%%%%%%%%%%%%%%%
%%%%%%%%%%%%%%%%%%%%%%%%%%%%%%%%%%%%
\bibliography{final}

\begin{thebibliography}{9}
\expandafter\ifx\csname natexlab\endcsname\relax\def\natexlab#1{#1}\fi
\expandafter\ifx\csname bibnamefont\endcsname\relax
  \def\bibnamefont#1{#1}\fi
\expandafter\ifx\csname bibfnamefont\endcsname\relax
  \def\bibfnamefont#1{#1}\fi
\expandafter\ifx\csname citenamefont\endcsname\relax
  \def\citenamefont#1{#1}\fi
\expandafter\ifx\csname url\endcsname\relax
  \def\url#1{\texttt{#1}}\fi
\expandafter\ifx\csname urlprefix\endcsname\relax\def\urlprefix{URL }\fi
\providecommand{\bibinfo}[2]{#2}
\providecommand{\eprint}[2][]{\url{#2}}

\bibitem[{\citenamefont{Pirner and Galow}(2009)}]{Pirner:2009gr}
\bibinfo{author}{\bibfnamefont{H.~J.} \bibnamefont{Pirner}} \bibnamefont{and}
  \bibinfo{author}{\bibfnamefont{B.}~\bibnamefont{Galow}},
  \bibinfo{journal}{Phys. Lett.} \textbf{\bibinfo{volume}{B679}},
  \bibinfo{pages}{51} (\bibinfo{year}{2009}), \eprint{0903.2701}.

\bibitem[{\citenamefont{McLerran and Svetitsky}(1981)}]{McLerran:1981pb}
\bibinfo{author}{\bibfnamefont{L.~D.} \bibnamefont{McLerran}} \bibnamefont{and}
  \bibinfo{author}{\bibfnamefont{B.}~\bibnamefont{Svetitsky}},
  \bibinfo{journal}{Phys. Rev.} \textbf{\bibinfo{volume}{D24}},
  \bibinfo{pages}{450} (\bibinfo{year}{1981}).

\bibitem[{\citenamefont{Witten}(1998)}]{Witten:1998qj}
\bibinfo{author}{\bibfnamefont{E.}~\bibnamefont{Witten}},
  \bibinfo{journal}{Adv. Theor. Math. Phys.} \textbf{\bibinfo{volume}{2}},
  \bibinfo{pages}{253} (\bibinfo{year}{1998}), \eprint{hep-th/9802150}.

\bibitem[{\citenamefont{Nian and Pirner}(2009)}]{Nian:2009mw}
\bibinfo{author}{\bibfnamefont{J.}~\bibnamefont{Nian}} \bibnamefont{and}
  \bibinfo{author}{\bibfnamefont{H.~J.} \bibnamefont{Pirner}}
  (\bibinfo{year}{2009}), \eprint{0908.1330}.

\bibitem[{\citenamefont{Galow et~al.}(2009)\citenamefont{Galow, Megias, Nian,
  and Pirner}}]{Galow:2009kw}
\bibinfo{author}{\bibfnamefont{B.}~\bibnamefont{Galow}},
  \bibinfo{author}{\bibfnamefont{E.}~\bibnamefont{Megias}},
  \bibinfo{author}{\bibfnamefont{J.}~\bibnamefont{Nian}}, \bibnamefont{and}
  \bibinfo{author}{\bibfnamefont{H.~J.} \bibnamefont{Pirner}}
  (\bibinfo{year}{2009}), \eprint{0911.0627}.

\bibitem[{\citenamefont{Meyer-Ortmanns}(1996)}]{MeyerOrtmanns:1996ea}
\bibinfo{author}{\bibfnamefont{H.}~\bibnamefont{Meyer-Ortmanns}},
  \bibinfo{journal}{Rev. Mod. Phys.} \textbf{\bibinfo{volume}{68}},
  \bibinfo{pages}{473} (\bibinfo{year}{1996}), \eprint{hep-lat/9608098}.

\bibitem[{\citenamefont{Yagi et~al.}(2005)\citenamefont{Yagi, Hatsuda, and
  Miake}}]{Yagi:2005yb}
\bibinfo{author}{\bibfnamefont{K.}~\bibnamefont{Yagi}},
  \bibinfo{author}{\bibfnamefont{T.}~\bibnamefont{Hatsuda}}, \bibnamefont{and}
  \bibinfo{author}{\bibfnamefont{Y.}~\bibnamefont{Miake}},
  \bibinfo{journal}{Camb. Monogr. Part. Phys. Nucl. Phys. Cosmol.}
  \textbf{\bibinfo{volume}{23}}, \bibinfo{pages}{1} (\bibinfo{year}{2005}).

\bibitem[{\citenamefont{Shoshi et~al.}(2002)\citenamefont{Shoshi, Steffen, and
  Pirner}}]{Shoshi:2002in}
\bibinfo{author}{\bibfnamefont{A.~I.} \bibnamefont{Shoshi}},
  \bibinfo{author}{\bibfnamefont{F.~D.} \bibnamefont{Steffen}},
  \bibnamefont{and} \bibinfo{author}{\bibfnamefont{H.~J.}
  \bibnamefont{Pirner}}, \bibinfo{journal}{Nucl. Phys.}
  \textbf{\bibinfo{volume}{A709}}, \bibinfo{pages}{131} (\bibinfo{year}{2002}),
  \eprint{hep-ph/0202012}.

\bibitem[{\citenamefont{Shoshi et~al.}(2003)\citenamefont{Shoshi, Steffen,
  Dosch, and Pirner}}]{Shoshi:2002rd}
\bibinfo{author}{\bibfnamefont{A.~I.} \bibnamefont{Shoshi}},
  \bibinfo{author}{\bibfnamefont{F.~D.} \bibnamefont{Steffen}},
  \bibinfo{author}{\bibfnamefont{H.~G.} \bibnamefont{Dosch}}, \bibnamefont{and}
  \bibinfo{author}{\bibfnamefont{H.~J.} \bibnamefont{Pirner}},
  \bibinfo{journal}{Phys. Rev.} \textbf{\bibinfo{volume}{D68}},
  \bibinfo{pages}{074004} (\bibinfo{year}{2003}), \eprint{hep-ph/0211287}.

\end{thebibliography}

\end{document}